\documentclass[aps,groupedaddress,prl]{revtex4}
\usepackage{graphicx}

\def\dc{{\cal D}_c}
\def\ds{{\cal D}_s}
\begin{document}

\title{Structural and computational depth of diffusion limited aggregation }
\author {D. Tillberg}
\author {J. Machta}
\email{machta@physics.umass.edu}
\affiliation{Department of Physics,
University of Massachusetts, Amherst, Massachusetts 01003}
\date{\today}

\begin{abstract}   
Diffusion limited aggregation is studied from the perspective of computational complexity.  A parallel algorithm is exhibited that requires a number of steps that scales as the depth of the tree defined by the cluster.  The existence of this algorithm suggests a connection between a fundamental computational and structural property of DLA.
\end{abstract}

\maketitle

Diffusion limited aggregation (DLA), introduced some 20 years ago by Witten and Sander~\cite{WiSa}, is a model of pattern formation and an example of self-organized criticality.  The dynamical rules for DLA are simple but the patterns produced are complex and have thus far defied full theoretical understanding though there has been recent progress for two-dimensional DLA~\cite{BaDaPr02,BaSo03}.  DLA models a number of physical systems including electrodeposition, fluid flow in porous media and the growth of bacterial colonies~\cite{Vi}.

The stochastic growth rules for DLA can be couched in terms of random walk dynamics.  In the present work we consider two-dimensional diffusion limited aggregates composed of circular particles of unit diameter.  The initial condition is a single seed particle at the origin.  The growth of the aggregate occurs one particle at a time and proceeds by starting the first particle at a random position on the ``birth circle,'' $r_1(0)=r_b>1$.  This particle does a random walk, ${\bf r}_1(t)$, until it either drifts out to the ``death circle,'' $r_1>r_d \gg r_b$, or comes in contact with the particle at the origin, $r_1=1$. If it contacts the particle at the origin it sticks and the aggregate grows.  If it reaches the death circle it is reborn at a random point on the birth circle and the process is repeated until the particle sticks.  After $n$ steps of this construction, the aggregate consists of $n$ connected particles just contained in a circle of radius $R_n$.  On the $n^{\rm th}$ step, a particle is launched on the birth circle, $r_b > R_n$, and follows a random walk ${\bf r}_n(t)$ until it sticks on one of the existing particles in the aggregate at sticking position ${\bf r}^\ast_n$ such that there is a $k<n$ and  $|{\bf r}^\ast_n-{\bf r}^\ast_k|=1$.  

\begin{figure}
\includegraphics[width=7in]{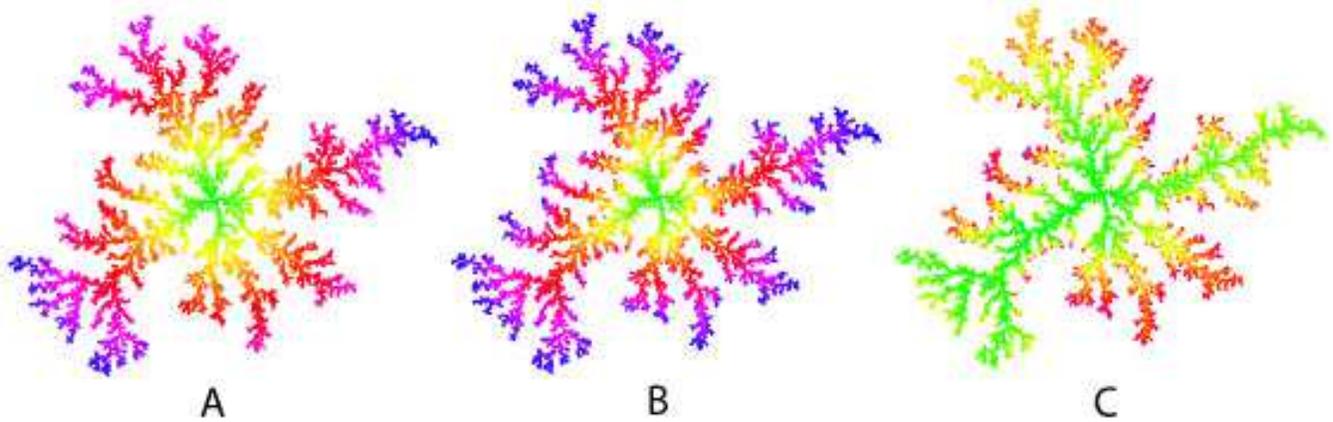}
\caption{Color-coded pictures of the same cluster: A, the chemical distance to the origin; B, the parallel step on which the particle first becomes semi-secure at the correct position; C, the deficit. Green (light) represents the lowest values and blue (dark) the highest.}
\label{fig:tryptich}
\end{figure}

DLA is believed to form fractal clusters such that $R_N\sim N^{1/d_f}$ where $R_N$ is the average radius of a cluster of $N$ particles and $d_f$ is the fractal dimension, estimated by numerical simulation to be $1.715 \pm .004$ ~\cite{ToMe}.  DLA clusters define a graph where the nodes are the particles and the edges are the contacts between particles.  Since each particle sticks to a single predecessor, this graph is a tree rooted at the origin.  One of the quantities that we will be interested in is the {\em structural depth}, $\ds$ of this tree, defined as the length of the path from the root to the outermost leaf.  Numerical simulations~\cite{MeMaHaSt84} support the conjecture that the structural depth grows linearly in the radius of the cluster, $\ds \sim R$.  The growth mechanism of DLA strongly favors adding particles at the outermost tips of the cluster and thus exerts a tension that tends to make the growth radial.  Figure \ref{fig:tryptich}A color codes particles in a cluster according to their distance to the seed particle along the tree, sometimes referred to as the chemical distance to the origin.  This figure reveals the close relation between chemical distance and Euclidean distance from the origin.

In addition to its structural properties DLA has computational properties that help elucidate its complexity.  The growth rules for DLA and the resulting patterns suggest that DLA has a great deal of {\em history dependence}.  The construction of a DLA cluster requires a long sequence of steps and some random choices made early in the growth process are frozen in and  have a large impact on the structure of the aggregate thereafter.  Indeed the stochastic history dependence present in DLA can be considered a marker for complexity.  Biological evolution,  the ultimate process for generating complexity, also has the feature that the present state is emerges from a long sequence of stochastic steps and that accidents occurring in early epochs are frozen in and create the ground rules for later epochs.

The intuitive notion of history dependence in stochastic models such as DLA can be formalized in the context of computational complexity theory.  The idea that the length of a history can be measured in the framework of computation was introduced by Bennett~\cite{Benn88, Benn90} though our treatment differs by emphasizing parallel rather than sequential computation and ensembles rather than individuals.  We define the {\em computational depth}, $\dc$ of a statistical physics models such as DLA as the minimum number of parallel computational steps needed to generate a typical system state of the model.  Both statistical physics and computational complexity theory are concerned with scaling properties, for example, in the case of DLA, we are interested in the leading large $N$ behavior of $R$, $\ds$ and $\dc$.

The notion of computational depth requires that we specify the  parallel computer used to simulate the system.  Computational complexity theory~\cite{Papa, GrHoRu} provides a standard, idealized model of parallel computation called the parallel random access machine or PRAM.   The PRAM consists of many simple processors all connected to a single global memory.  All processor run the same program synchronously though each processor has a distinct label so that the program can directs different processors to do different calculations.  In a single computational step, each processor carries out an elementary logical or arithmetic operation and exchanges information with the global memory.  Since we are considering a stochastic system random numbers are needed and we assume a sufficient supply are stored in the global memory before the simulation begins.  The PRAM is a massively parallel model of computation where the number of processors and memory cells is allowed to grow {\em polynomially} (i.e. as a power) in the size of the system, in our case the number of particles $N$. Since any processor in the PRAM can communicate with any memory element in a single time step, the PRAM is not a realistic, scalable model of parallel computation and the considerations of this paper are not intended to provide a practical means of simulating DLA. Instead, the goal is to elucidate whether assembling a DLA cluster requires a long sequence of steps.  The PRAM is one of many equivalent models of parallel computation and the scaling behavior for $\dc$ reflects an intrinsic property of DLA rather than a particular strength or weakness of the PRAM.  With this preamble we can now define computational depth as the average number of steps needed to construct a system state on a PRAM using the fastest parallel algorithm. 

The primary objective of this paper is to show that $\dc \lesssim \ds$ and to motivate the conjecture that in fact $\dc \sim \ds$~\footnote{The symbols $\lesssim$ and $\sim$ are to be interpreted, respectively, as asymptotic inequality and equality  up to logarithmic factors, e.g. $N^3 \log^2 N \lesssim N^3$.}.  Previous work has placed upper bounds on the computational depth of various pattern formation processes in statistical physics.  For example, it has been shown that the clusters formed by the Bak-Sneppen model~\cite{MaLi01}, internal DLA~\cite{MoMa00}, invasion percolation, the Eden model and ballistic deposition~\cite{MaGr} all have {\em polylog} depth $\dc \lesssim \log^{{\cal O}(1)} N$, where $N$ is the number of degrees of freedom of the system.  These results show that these models do not have a strong intrinsic history dependence since constructing typical states can be carried out in a small number of parallel steps. These models generate trees whose depth greatly exceeds the computational depth, $\dc \ll \ds$.   The situation is more complicated for DLA.  It has been shown~\cite{MaGr96} that random walk dynamics for DLA defines a {\bf P}-complete problem, which strongly suggests that there is no polylog depth parallel construction for DLA, though it does not rule out a better power law than the linear scaling with $N$ of the conventional, one walk at a time growth rule.  The question posed here is what is the best power law scaling for a parallel simulation of DLA.  Previous work~\cite{MoMaGr97, BaDaPr02} showed that $\dc \lesssim N^{0.74} \sim R^{1.26}$.  Here we improve that bound by exhibiting a faster parallel algorithm than the one described in~\cite{MoMaGr97}.

The parallel algorithm for DLA proposed here assembles the cluster iteratively.  On each step, every particle is moved to a tentative position so that the true cluster is approximated with increasing fidelity.  Before the iteration is begun, an ordered list of $N$ sufficiently long random walk trajectories $\{{\bf r}_i(t)|i=1,\ldots,N\}$ are generated and stored in memory~\footnote{Random walk trajectories may include several restarts from the birth circle each time the death circle is reached.}.  The final cluster generated by the parallel algorithm is the same as the cluster that would result using these random walks trajectories and the standard sequential rules for DLA.

In each step, all particles are moved along their trajectories to tentative sticking points defined by a temporary cluster called  the {\em semi-secure} cluster.  On the first step, the semi-secure cluster consists only of the seed particle and each particle sticks at the point where it first contacts the seed particle, ignoring overlaps with other particles.  The semi-secure cluster for the second step consists of the seed particle and any particle whose path to the seed did not cross the sticking point of any lower-numbered particle.  On the second step every particle is moved independently along its trajectory until it first contacts this semi-secure cluster.  Note that it is possible that a given particle on a given step may be in the semi-secure cluster in one location but stick in a different location.

At the beginning of the $m^{\rm th}$ step in the algorithm, we have a semi-secure cluster defined by $N^{(m)}$ particles at locations $\{{\bf s}^{(m)}_j | j \in S^{(m)}\}$ where $S^{(m)}$ is the set of semi-secure particle indices at step $m$.  Each particle is then moved to a location based on the template provided by this semi-secure cluster. The sticking point of a particle is determined by independently moving it along its trajectory until it first contacts a semi-secure particle location with an index lower than its own.  ${\bf r}^{(m)}_i$ is the sticking point of particle $i$ at step $m$ if it is the first time $\tau_i^m$ along the trajectory such that there is contact with a semi-secure particle $j<i$ and $|{\bf r}_i(\tau_i^m)-{\bf s}^{(m)}_j|=1$.

After particles are moved to their sticking points the current semi-secure cluster is discarded and the locations of the sticking points used to determine the next semi-secure cluster.  Each particle may be categorized as ``semi-secure" or ``not semi-secure.''  A particle $i$ is semi-secure at the beginning of step $m+1$ if the path to its sticking point in step $m$ does not intersect the sticking point of any predecessor particle on step $m$.  More formally, $i \in S^{(m+1)}$ and ${\bf s}^{(m+1)}_i={\bf r}^{(m)}_i$, if for all $j<i$ and all $t<\tau_i^m$,  $|{\bf r}_i(t)-{\bf r}_j^{(m)}|>1$.  A semi-secure particle such that all its predecessors are also semi-secure is called {\em secure}.  It is easy to see that a secure particle is in the correct location in the cluster.  The parallel algorithm has constructed a correct DLA cluster of size $M$ for the given trajectories as soon as particle $M$ is secure.  

There is one additional rule for constructing the semi-secure cluster.  It may happen that the semi-secure cluster defined by the above rules is a multiply connected ``forest" composed of several ``trees."  In this case, only the tree rooted at the seed particle is retained and the remaining particles, not connected to the seed, are removed from the semi-secure cluster.

As discussed in detail in~\cite{MoMaGr97}, random walk trajectories can be generated, sticking points of these trajectories to an existing cluster found and interferences among particles identified in polylog parallel time using polynomially many processors.  Using a  parallel algorithms for connected components~\cite{GiRy}, one can identify the the tree connected to the origin in polylog time.  Thus, the setup stage and each step of the algorithm require polylog time so that, up to logarithmic factors, the parallel  time used by the algorithm scales as the number of steps.

\begin{figure}
\includegraphics[width=3.5in]{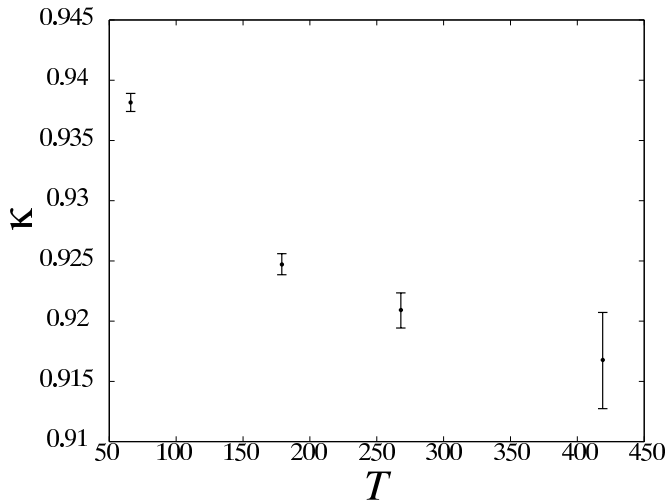}
\caption{The ratio $\kappa=\ds/T$ of the average structural depth of the cluster divided by the number of parallel computational steps $T$  vs.\ $T$. }
\label{fig:ratio}
\end{figure}

How well does this algorithm perform?  It is easy to see that at least one new particle becomes secure on each iteration so that the parallel algorithm requires no more than $N$ iterations.  In fact, the performance is much better than this weak bound.  We have simulated the parallel algorithm on a sequential computer for $N$ up to 20,000.  Let $T$ be the number of iterations required by the algorithm and let $\kappa=\ds/T$ be the ratio of the number of iterations divided by the structural depth. Figure \ref{fig:ratio} shows $\kappa=\ds/T$ plotted against $T$.  The average is over several thousand clusters for the small sizes and 63 cluster for $N=20,000$.  Although $\kappa$ decreases slightly as $T$ and $N$ increase, the data strongly suggests that $\kappa$ approaches an asymptote near $0.9$.  Thus, up to logarithmic factors, the computational depth of DLA is no greater than the structural depth.  In terms of a dynamic exponent for the algorithm, defined by $T \sim R^z$, the data suggests that $z = 1$.

The algorithm of \cite{MoMaGr97} is similar to the present algorithm except that the template used in step $m+1$ is the secure cluster from step $m$ rather than the semi-secure cluster.  The semi-secure cluster is larger than the secure cluster but, unlike the secure cluster, it contains particles at incorrect positions which must later be moved.  Our numerical results show that the larger size of the semi-secure cluster more than compensates for the errors in the semi-secure cluster.  The dynamic exponent of the  ``secure'' algorithm of \cite{MoMaGr97} is $1.26$ compared to $1$ for the ``semi-secure'' algorithm described here.

\begin{figure}
\includegraphics[width=6in]{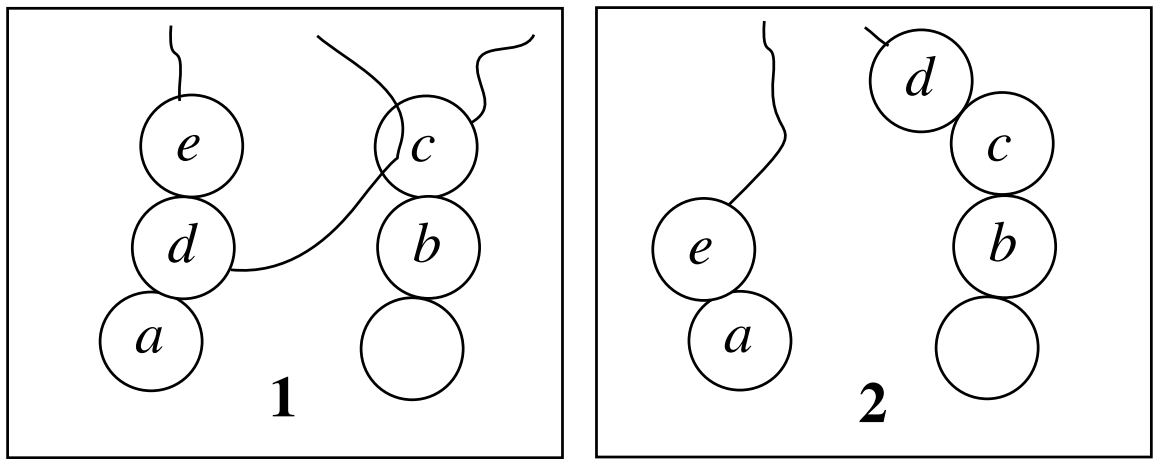}
\caption{A sequence of two parallel steps that increases the deficit of particle. Suppose that at the outset of the step shown in panel 1, all particles but $c$ are semi-secure--$b$ has just become semi-secure and $e$ has been semi-secure in its location for the past $k$ steps. After step 1,  $d$ and $e$ are no longer semi-secure, $d$ due to interference with $c$ and $e$ because it is no longer connected to the origin.  In step 2, $e$ and $d$ become semi-secure in new locations and $e$'s deficit is increased by $k+2$. }
\label{fig:fall}
\end{figure}

If $\kappa$ is indeed asymptotically near one, it implies that on most steps of the parallel algorithm, one new level is added to the tree defined by the cluster. It is obvious that the algorithm cannot add more than one level to the tree in one parallel step so $\kappa \leq  1$.  It is instructive to consider the way that $T$ becomes larger than the structural depth.  After each parallel step, every semi-secure  particle can be assigned a {\em deficit}.  A particle's deficit is defined as the difference between the step on which it most recently became semi-secure at it's current location and its chemical distance from the origin.  Figure \ref{fig:fall} shows how the deficit of a particle can increase by an arbitrary amount. The two panels show two successive parallel steps.  Particle $b$ has just become semi-secure in the step preceding panel 1.   Trajectories and sticking points for particles $c$, $d$ and $e$ are shown.  The ordering for the particles is assumed to be $b<c<d<e$ and $a<d$.  Before step 1, all particles except $c$ are semi-secure.   Suppose that $e$ had become semi-secure at its current location $k$ steps before step 1. In panel 1, particle $c$ becomes semi-secure  and interferes with $d$ so $d$ is no longer semi-secure.  In panel 2, both $d$ and $e$ become semi-secure in new positions.  The deficit of $e$ is increased by $k+2$ over what it was before panel 1 since its chemical distance is decreased by one and it has become semi-secure $k+1$ steps later.  When the algorithm is finished and all particles are secure, deficits along branches of the tree are a non-decreasing function of chemical distance. The maximum, taken over leaves of the tree, of the sum of the deficit and chemical distance to the origin gives the running time of the algorithm in parallel steps.

Figure \ref{fig:tryptich}  shows three images of the same cluster colored to reveal different properties.  This cluster consists of 20,000 particles, it has a structural depth  of 439 and requires 479 steps to assemble.    In \ref{fig:tryptich}A the particles are colored according to their structural depth (chemical distance from the origin).  Note that the contours of equal structural depth are nearly circular.  In \ref{fig:tryptich}B particles are colored according to the parallel step on which they first become semi-secure at their final locations.  This image reveals that growth generated by the parallel algorithm conforms more to the shape of the branches of the cluster. In \ref{fig:tryptich}C particles are colored according to their deficits, \ref{fig:tryptich}C shows the difference between the quantities in \ref{fig:tryptich}B and \ref{fig:tryptich}A and it reveals that the dominant branches have small deficits but that large deficits can develop in branches which are less robust and screened by the dominant branches.  The reason for this tendency can be seen in the example of Fig.\ \ref{fig:fall} where a theft from one branch to another increases the deficit on the branch that loses the particle. 

The algorithm proposed here assembles DLA clusters by adding nearly one level of structural depth in each parallel step.  Our intuition is that it is not possible to generate DLA clusters in substantially fewer steps than the structural depth of the cluster so that $\dc \sim \ds$. This conjecture links fundamental structural and computational properties of DLA. Like other lower bounds in computational complexity theory, this conjecture is likely to be difficult to prove.  
 
\acknowledgements
This work was supported by NSF grants DMR-9978233 and DMR-0242402.

\end{document}